\documentclass[sigconf,screen]{acmart}
\usepackage{microtype}
\usepackage{xspace}
\usepackage{tikz}
\usetikzlibrary{positioning}

\usepackage{makecell} 
\usepackage{multirow}
\usepackage{algorithm}
\usepackage{algorithmic}
\usepackage{amsfonts}
\usepackage{xcolor}
\usepackage{booktabs}
\usepackage[most]{tcolorbox}
\usepackage{calligra}
\newcommand{\std}[1]{\scriptsize$\pm$#1}

\newcommand{\frameworkname}{ExeCRE}

\newcommand{\vtsubname}{Voting}

\newcommand{\framework}{\frameworkname\xspace}
\newcommand{\frameworksub}{\frameworkname}

\newcommand{\vtsub}{\frameworksub-\vtsubname\xspace}

\usepackage{fancyvrb}
\usepackage{fvextra}

\usepackage{listings}
\newtcolorbox{promptbox}{
  colback=gray!10!white, colframe=black, sharp corners,
  boxrule=0.3mm, top=2pt, bottom=2pt, left=2pt, right=2pt, breakable,
  fontupper=\footnotesize
}

\setcopyright{cc}
\setcctype{by}
\acmDOI{10.1145/3832783.3837445}
\acmYear{2026}
\copyrightyear{2026}
\acmISBN{979-8-4007-2882-2/2026/10}
\acmConference[ASE '26]{Proceedings of the 41st IEEE/ACM International Conference on Automated Software Engineering}{October 12--16, 2026}{Munich, Germany}
\acmBooktitle{Proceedings of the 41st IEEE/ACM International Conference on Automated Software Engineering (ASE '26), October 12--16, 2026, Munich, Germany}
\acmSubmissionID{ase26main-p496-p}
\received{2026-03-27}
\received[accepted]{2026-06-18}

\begin{document}

\title{ExeCRE: Execution-Consistency Guided Reliability Estimation for Self-Correcting Code Generation}

\author{Yiru Dong}
\orcid{0009-0002-1733-0391}
\affiliation{%
  \institution{Beihang University}
  \city{Beijing}
  \country{China}
}
\email{yrdong@buaa.edu.cn}

\author{Richong Zhang}
\orcid{0000-0002-1207-0300}
\affiliation{%
  \institution{Beihang University}
  \city{Beijing}
  \country{China}
}
\email{zhangrichong@buaa.edu.cn}

\author{Fanshuang Kong}
\correspondingauthor
\orcid{0000-0002-9046-740X}
\affiliation{%
  \institution{Beihang University}
  \city{Beijing}
  \country{China}
}
\email{kongfs@buaa.edu.cn}

\author{Si Chen}
\orcid{0009-0009-5842-8898}
\affiliation{%
  \institution{Beihang University}
  \city{Beijing}
  \country{China}
}
\email{chen.si@buaa.edu.cn}

\begin{abstract}
Large language models (LLMs) have made notable progress in code generation, but they still struggle on challenging tasks that require sophisticated algorithms or complex implementations. Recent methods increasingly use code execution as feedback, especially in self-correction pipelines that construct verification signals from generated code. However, these pipelines often depend on supervision signals whose reliability is unknown, which can introduce misleading feedback, unnecessary revisions, and incorrect final answers.
To address this issue, we propose ExeCRE, an \textbf{Exe}cution-\textbf{C}onsistency guided code \textbf{R}eliability \textbf{E}stimation framework. Instead of judging candidate code by tests or LLM feedback, ExeCRE estimates code reliability by statistically analyzing consistency patterns in execution outputs over a large number of randomly generated inputs. It collects execution outputs over generated inputs, projects them into consistency signals, and applies Dawid--Skene model to infer latent code reliability.
We integrate ExeCRE into self-correction for code generation. Experiments show that ExeCRE consistently improves both effectiveness and stability, while substantially reducing misleading correction signals. Under GPT-5.2 on LiveCodeBench, the average number of misleading feedback cases on already correct code drops from 113.2 with a representative self-correction baseline to 14.0 with ExeCRE.
As an additional study, we apply the same reliability estimation strategy to code-based mathematical reasoning and observe similar benefits.
These results suggest that ExeCRE enables more reliable use of generated code in execution-based pipelines.
\end{abstract}


\maketitle

\section{Introduction}

Large language models (LLMs) have achieved remarkable success in code generation, particularly on elementary programming tasks~\cite{lozhkov2024starcoder,iyer2023humanevalplus,liu2025deepseek,10.1145/3641289,guo2024deepseekcoder}. However, their performance degrades on more challenging problems that require sophisticated algorithms or complex implementation details~\cite{hendrycksmeasuring,jimenezswe,zhang2024livecodebench,gehring2024rlef}. To mitigate this limitation, a prominent and increasingly effective research paradigm is self-correction, where models iteratively refine generated code by leveraging execution-based feedback signals~\cite{shinn2023reflexion,madaan2024selfrefine,chen2025revisit,yuksekgonul2025optimizing}.

A fundamental prerequisite for effective self-correction is reliable verification. These verification signals determine whether candidate solutions should be revised and therefore directly shape the refinement trajectory.
One common way to obtain such verification signals is through execution-based tests. Recent studies suggest that code execution often provides more informative feedback than LLM-based judgments~\cite{ding2024cycle,li2025codeprm,chen2025revisit,zhong2024debug}. They also show that reference codes, although often brute-force and inefficient, can still be used to construct such tests~\cite{zhang2023algo}.

\begin{figure}
    \centering
    \includegraphics[width=\columnwidth]{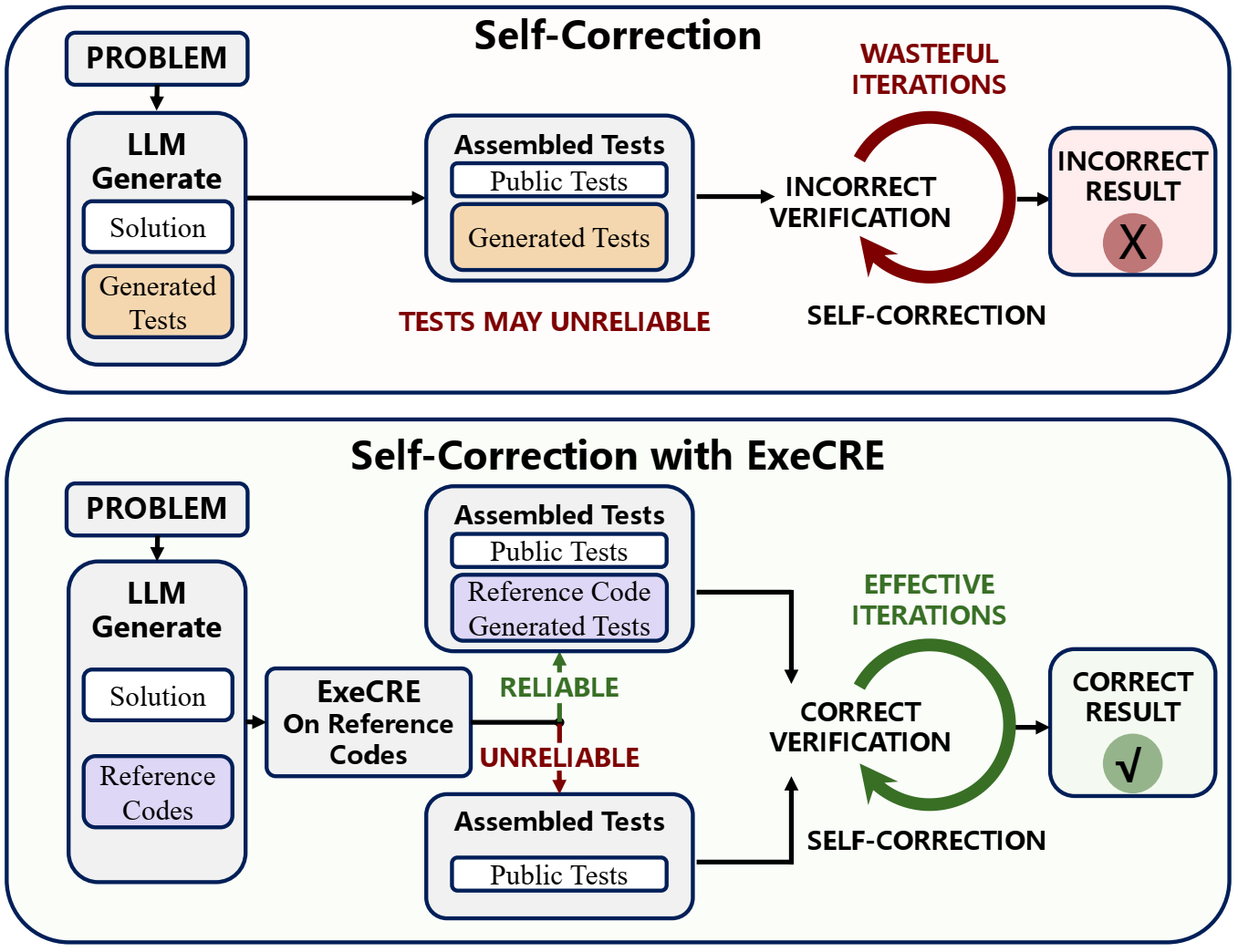}
    \caption{Self-correction pipelines with and without ExeCRE. ExeCRE filters unreliable reference codes before they produce misleading verification signals.}
    \label{fig:Self_Correction}
\end{figure}

However, in practice, reference codes themselves may be unreliable: they can contain subtle bugs, cover only partial cases, or disagree with each other on complex inputs~\cite{zhang2023algo}. Using unreliable reference codes to construct verification signals can mislead the self-correction process, leading to correct codes being unnecessarily revised or incorrect codes being reinforced.
To improve the reliability of LLM-generated results, prior work has explored consistency as an effective signal~\cite{chen2023codet,huang2024enhancing,sun2022recitation,wang2023selfconsistency,xiong2023examining,valentin2026incoherence}. In code generation, consistency has been used to favor solutions that pass more tests or exhibit more consistent behavior across different counterparts. Instead of relying on generated tests, our proposed ExeCRE directly analyzes large-scale execution outputs and uses a statistical method to identify unreliable reference codes. This reliability-aware design substantially reduces misleading supervision during self-correction.
On LiveCodeBench, ExeCRE substantially reduces misleading feedback on already correct code, as shown in Table~\ref{tab:misleading-feedback}.

As illustrated in Figure~\ref{fig:Self_Correction}, reference code reliability is a critical yet underexplored bottleneck in execution-guided self-correction. Existing approaches~\cite{zhang2023algo,yuksekgonul2025optimizing} typically assume reference codes or tests to be correct, or rely on language models to judge correctness, leaving the reliability of supervision signals largely unquantified.
Representative methods for self-correction and verification differ substantially in how feedback is obtained, how correctness is judged, and what information is used to construct such signals. As summarized in Table~\ref{tab:method_compare}, existing approaches either rely on language models as judges, use relatively small numbers of generated tests, or assume the reliability of reference codes without explicitly estimating it. This design space motivates a more direct treatment of reference code reliability.

\begin{table}[ht]
\centering
\small
\setlength{\tabcolsep}{4pt}
\caption{Comparison of representative self-correction and verification approaches.}
\label{tab:method_compare}
\begin{tabular}{lcccc}
\toprule
Method & \begin{tabular}[c]{@{}c@{}}Feedback\\Source\end{tabular} & 
\begin{tabular}[c]{@{}c@{}}Judging\\Basis\end{tabular} &
\begin{tabular}[c]{@{}c@{}}Information\\Source\end{tabular} & Scale \\
\midrule
ConTested & Test & LLM/human & LLM-gen. tests & $\sim$10 \\
Oracle-Guided    & LLM  & LLM       & Symbolic inputs & $\sim$1000 \\
B4        & Test & Statistical & LLM-gen. tests & $\sim$100 \\
ALGO      & Test & Ref. code   & LLM rand. inputs & $\sim$30 \\
TextGrad  & LLM  & LLM       & -- & -- \\
\midrule
ExeCRE    & Test & Statistical & Parsed rand. inputs & $\sim$1000 \\
\bottomrule
\end{tabular}
\end{table}

To address this issue, we propose an \textbf{Exe}cution-\textbf{C}onsistency guided reference code \textbf{R}eliability \textbf{E}stimation framework, named ExeCRE, which explicitly models and infers the reliability of reference codes and leverages only trustworthy supervision for self-correction.
Specifically, ExeCRE operates in three stages. First, we introduce a schema-driven test input generation mechanism that enforces structural and semantic constraints, enabling the automatic generation of a large number of valid and diverse test inputs. Second, motivated by prior observations that brute-force implementations, while inefficient, tend to be highly accurate~\cite{zhang2023algo}, we execute multiple candidate reference codes on these inputs and analyze their execution consistency. We then employ Dawid–Skene model~\cite{dawid1979maximum} to infer the latent error-rates of reference codes from their consistency patterns, without requiring ground-truth labels. Finally, we select the reference code with the highest inferred reliability and use the tests constructed from it as trusted execution judges to guide iterative self-correction.
By explicitly filtering unreliable supervision signals, ExeCRE reduces unnecessary or harmful revisions, avoids cascading errors during refinement, and enables more stable and effective code generation. 

We summarize our contributions as follows:
\begin{itemize}
    \item We propose \framework, an \textbf{Exe}cution-\textbf{C}onsistency guided reference code \textbf{R}eliability \textbf{E}stimation framework for modeling the reliability of codes from execution outputs.
    
    \item We integrate ExeCRE into execution-guided self-correction by selecting or filtering reference codes before they are used to construct verification signals, thereby reducing the risk of misleading supervision.
    
    \item Experiments on 182 LiveCodeBench problems, together with a small applicability study on GSM8K, show that ExeCRE improves self-correction effectiveness and stability while substantially reducing unnecessary correction signals.
\end{itemize}

\section{Preliminary}
\label{sec:preliminary}
Given a problem instance \(P\), ExeCRE collects multiple candidate codes for the same problem and analyzes their execution outputs on a set of constructed inputs.
Formally, let \(C_P = \{C_1, C_2, \dots, C_{|C_P|}\}\) denote the set of candidate codes generated for \(P\), and let \(I_P = \{I_1, I_2, \dots, I_{|I_P|}\}\) denote the set of execution inputs.
Running candidate code \(C_j\) on input \(I_i\) produces an execution output \(O_{ij}\), and all outputs form an execution matrix \(O = [O_{ij}]\).

Our goal is not to formally verify full semantic correctness, but to estimate whether a candidate code is reliable enough for subsequent use.
The intuition is that more reliable codes tend to exhibit more stable agreement patterns across diverse inputs, while unreliable ones are more likely to behave inconsistently.
We refer to this signal as \textit{execution consistency}. Based on this signal, ExeCRE assigns each candidate code \(C_j\) a reliability score \(\alpha_j \in [0,1]\).

Our formulation is related to Dawid--Skene (DS), a classical EM-based method for aggregating noisy annotations.
In the standard DS setting, multiple annotators provide labels for the same item, and the model estimates both latent true labels and annotator-specific error-rates.
In our setting, execution inputs correspond to items, candidate codes correspond to annotators, and the observed signals are derived from execution consistency rather than raw execution outputs.
This view allows us to estimate the reliability of code from execution behavior.

\section{ExeCRE}
\label{sec:method}

ExeCRE estimates the reliability of candidate reference codes before they are used to generate verification signals. Given multiple reference code candidates for the same problem, it constructs diverse execution inputs, collects execution outputs, projects them into binary consistency signals, and applies a Dawid--Skene based estimator to infer candidate reliability. The resulting scores are used to either select a trusted reference code or trigger fallback when no candidate is sufficiently reliable.

\begin{figure*}[t]
    \centering
    \includegraphics[width=\textwidth]{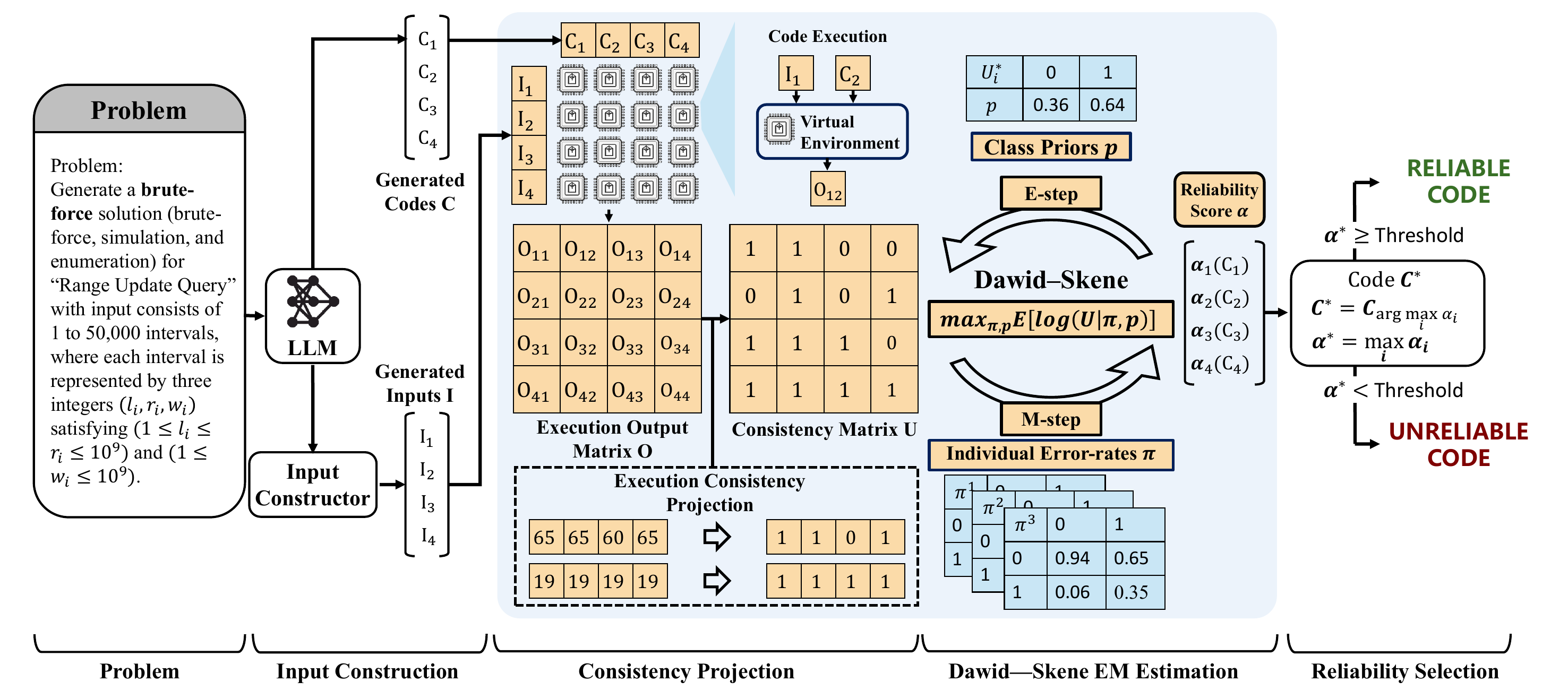}
    \caption{Overview of ExeCRE. Given a problem, ExeCRE constructs execution inputs and collects execution outputs from multiple candidate codes. It then projects raw outputs into a binary consistency matrix and applies Dawid--Skene EM to estimate the latent reliability of each candidate. The estimated reliability scores are finally used to select reliable code or trigger fallback when no candidate exceeds the threshold.}
    \label{fig:framework}
\end{figure*}

\subsection{Execution Input Construction for Code Generation}

ExeCRE constructs diverse execution inputs to expose behavioral differences among candidate codes. Unlike prior methods ~\cite{chen2023codet,dong2025contested,shinn2023reflexion} that rely on repeated LLM-based test generation or direct parameter parsing, ExeCRE prompts an LLM once to extract an input schema and then uses rule-based parsing to support both stdin-style and function-style inputs. This design reduces LLM cost while preserving broad input-format coverage and generating behaviorally discriminative inputs.

We prompt an LLM to extract a structured input schema from the problem description. The schema captures the input format, variable types, value ranges, and dependency relations among variables. The prompt specifies the target JSON format, explains the semantics of each field, and provides a few-shot examples to guide the extraction process. The full prompt is included in the publicly archived materials.
As illustrated in Listing~\ref{lst:schema_example}, input variables may exhibit explicit ordering constraints. For example, the schema may specify $r_i \in [l_i, 10^9]$, meaning that $l_i$ must be generated before $r_i$. To capture such relations, we extract a variable dependency graph from the schema, where each node corresponds to a variable and each directed edge $x \rightarrow y$ indicates that the valid range or constraints of $y$ depend on the value of $x$. The topological ordering of this graph then defines a valid generation order.

\begin{lstlisting}[caption={Example schema for structured input generation.}, label={lst:schema_example}]
{
  "parameters": {
    "intervals": {
      "type": "List",
      "range": [1, 50000],
      "data_type": {
        "l_i": {"type":  "int", "range":  [1, 1000000000]},
        "r_i": {"type":  "int", "range":  ["$l_i", 1000000000]},
        "weight_i": {"type":  "int", "range":  [1, 1000000000]}
      }
    }
  },
  "output_order" : [
    {"name":  "intervals", "separator":  "\n"}
  ]
}
\end{lstlisting}

Following this order, we sequentially instantiate variables by randomly sampling values within their specified types and constraint ranges, conditioned on previously instantiated variables. Symbolic bounds are resolved against the current variable context before sampling. For structured objects such as lists, groups, and matrices, the generator first resolves their size parameters and then recursively instantiates their elements. To keep generation valid and computationally manageable, the resolved ranges are further intersected with predefined runtime limits on numeric values and collection sizes. In our implementation, the sizes of List, Group, and Matrix objects are additionally bounded at 20, which reduces execution overhead while remaining sufficient to expose behavioral differences among functionally distinct code candidates.

To improve robustness, we further validate each schema candidate during execution. In rare cases, a schema may not be parsed or may produce inputs that trigger frequent runtime errors, typically due to mismatches in input format or value ranges. We therefore discard a schema candidate if it cannot be parsed or if more than 10\% of input--code executions fail under that schema. In implementation, we generate multiple schema candidates and keep only those that pass this validation step.
When a generated schema cannot be parsed or instantiated, our implementation retries with a new schema candidate for up to three attempts. Overall, this process yields a large and diverse set of valid execution inputs for subsequent reliability estimation. Formally, given a problem instance $P$, we construct a set of execution inputs $\mathcal{I}_P = \{ I_1, I_2, \dots \}$.

\subsection{Reliability Estimation from Execution Consistency}

To estimate the reliability of codes, we first construct an execution
output matrix $\mathbf{O}$. For a problem instance $P$, we generate a set of reference codes $\mathcal{C}_P = \{ C_1, C_2, \dots \}$ by prompting a large language model multiple times.
To expose subtle execution differences between candidate codes, we construct a large and diverse input set $\mathcal{I}_P$ based on the extracted schema. Correct and incorrect codes may behave identically on a few simple inputs, so limited testing often fails to expose their differences, which only emerge when the codes are executed on a broader and more diverse input set.
Collecting execution outputs for all pairs of reference codes and test inputs yields an execution output matrix, where $O_{ij}$ denotes the output of code $C_j$ on input $I_i$:
\[
\mathbf{O} = [O_{ij}]_{i=1,\dots,|\mathcal{I}_P|}^{j=1,\dots,|\mathcal{C}_P|}.
\]
For some input--code pairs, execution may fail, typically due to mismatches in input format or value ranges. In such cases, we record the corresponding entry $O_{ij}$ as $\mathrm{NaN}$ and exclude it from subsequent consistency projection and estimation. Since our goal is to identify correct code rather than distinguish fine-grained failure types, we do not model different error categories separately.

As a simple and intuitive instantiation for estimating code reliability, we
introduce \vtsub\ based on the execution output matrix $\mathbf{O}$. Specifically, reliability is computed as the average fraction of other candidate
codes that produce the same execution output as the target code across all
inputs, formulated as:
\[
\alpha^{\text{Voting}}_j
=
\frac{1}{|\mathcal{I}_P|}
\sum_{i=1}^{|\mathcal{I}_P|}
\frac{1}{|\mathcal{C}_P|}
\sum_{k=1}^{|\mathcal{C}_P|}
\mathbf{1}\!\left( O_{ik} = O_{ij} \right).
\]
However, this estimation treats all codes equally. When wrong codes fail on different subsets of inputs, simple voting can still be misled by the wrong codes that happen to agree on one input. Dawid--Skene instead models a latent label for each input and an error-rate for each code, then fits these quantities by maximizing the likelihood of the observed consistency labels. The resulting reliability scores can assign lower reliability to codes whose outputs fit the inferred labels poorly, rather than giving every candidate the same vote.
We therefore employ a Dawid–Skene-based latent reliability estimator that learns code-specific error tendencies from binary execution-consistency signals.

\subsubsection{\textbf{Dawid--Skene Model in Execution}}

In practice, codes exhibit distinct individual error characteristics: some codes are consistently correct across inputs, while others produce errors on a small fraction of inputs or fail consistently on certain input patterns.
Treating all codes as equally reliable will ignore such individual error characteristics and lead to suboptimal reliability estimates. 
To address this limitation, we employ an improved Dawid--Skene algorithm to infer the latent error-rates of each code from execution consistency signals.

The original Dawid--Skene algorithm is designed for annotation settings, where multiple annotators provide labels for the same instances.
Because these labels may be noisy, the model estimates both the latent true-label distribution and the annotator-specific error-rates under an iterative maximum-likelihood framework~\cite{dawid1979maximum}.
For annotator $j$, the error-rate matrix $\pi_{qa}^{(j)}$ models the conditional probability of producing observed label $a$ when the latent true label is $q$.
The key assumption is that annotators may exhibit systematic biases rather than purely random noise.
For example, an annotator may tend to confuse one class with another under certain latent labels, and such patterns can be estimated by aggregating observations across many instances.

\subsubsection{\textbf{Execution Consistency Projection}}
In the code generation task, however, when codes are treated as annotators and inputs as data instances, the effective annotation space corresponding to distinct execution outputs typically exceeds the number of observed inputs.
As a result, the Dawid–Skene model struggles to estimate code error-rates in this setting.

To address this issue, we reduce the annotation space by projecting raw execution outputs onto execution consistency signals.
Rather than modeling the concrete execution outputs directly, we focus on whether different codes exhibit consistent behavior on the same input.
Specifically, for each input--code pair, the execution output is mapped to a binary agreement indicator based on majority consistency across codes, where $\mathbf{1}[\cdot]$ denotes the indicator function that equals 1 when the condition holds and 0 otherwise:
\[
U_{ij}
=
\begin{cases}
\mathbf{1}\!\left[
O_{ij}
=
\arg\max_{v}\left|\{\, k \mid O_{ik}=v \,\}\right|
\right], & \text{if } O_{ij}\neq \mathrm{NaN},\\
\mathrm{NaN}, & \text{otherwise}.
\end{cases}
\]
This projection transforms the execution output matrix $\mathbf{O}$ into a consistency matrix $\mathbf{U}$, whose valid entries take values in $\{0,1\}$ and whose missing entries are recorded as $\mathrm{NaN}$. In the rare case of a tie, a majority value is arbitrarily selected.

This binary projection does not assume that the majority output is correct. Instead, each input is associated with a latent consistency state indicating which side of the partition better reflects reliable behavior. In particular, reliable behavior may correspond to the minority side when several codes share the same systematic mistake. This preserves agreement patterns while reducing the label space enough for Dawid--Skene estimation.
Under this view, different codes can still exhibit distinct conditional bias patterns on $\mathbf{U}$.
A more reliable code tends to agree with the latent consistency state more often across inputs, whereas a less reliable code may systematically fall on the opposite side for certain input patterns.
By collapsing diverse execution outputs into consistency and inconsistency, the projection preserves agreement patterns and systematic bias patterns among codes while substantially reducing the annotation space.
This makes code error-rate estimation with the Dawid--Skene algorithm tractable.

\subsubsection{\textbf{EM Formulation with Execution Consistency}}

After the execution-consistency projection, each input $I_i$ is associated with a binary observation from each candidate code $C_j$, namely $U_{ij}\in\{0,1\}$.
Here, the latent class does not represent a concrete execution output.
Instead, it represents the latent consistency state of the input, namely, which side of the observed binary partition is more likely to correspond to the reliable behavior for that input. We implement the Dawid--Skene estimation using Crowd-Kit~\cite{ustalov2024crowdkit}.
Following Dawid--Skene~\cite{dawid1979maximum},  we introduce for each input $I_i$ a pair of latent indicator variables
$\{T_{iq}\}_{q\in\{0,1\}}$,
where $T_{iq}=1$ if and only if the latent consistency class of $I_i$ is $q$, with
$\sum_{q\in\{0,1\}} T_{iq}=1$.
Let $p_q=P(T_{iq}=1)$ denote the class prior, and let
\[
\pi^{(j)}_{qa}=P(U_{ij}=a\mid T_{iq}=1)
\]
denote the individual error-rate parameters of candidate code $C_j$, namely, the probability that $C_j$ assigns observed label $a$ when the true latent class is $q$, where $q,a\in\{0,1\}$.
In this form, $\pi^{(j)}$ captures whether code $C_j$ tends to align with or deviate from the latent consistency state under different inputs, thereby modeling code-specific systematic tendencies on the binary consistency signals rather than on the raw outputs themselves.
For notational consistency with the original Dawid--Skene formulation, we further define
\[
n^{(j)}_{ia} = \mathbf{1}\{U_{ij}=a\}, \qquad a\in\{0,1\}.
\]

Here, $n^{(j)}_{ia}$ denotes the count indicator for whether candidate code $C_j$ produces observed label $a$ on input $I_i$. If the observation is missing, we set $n^{(j)}_{ia}=0$ for both $a=0$ and $a=1$. Let $\Omega_i = \{\, j \mid U_{ij}\neq \mathrm{NaN} \,\}$ denote the set of candidate codes whose consistency labels are available on input $I_i$.
In our setting, observations associated with Runtime Error or Time Limit Exceeded are mapped to $\mathrm{NaN}$ and are therefore excluded from $\Omega_i$.

Under the Dawid--Skene conditional independence assumption, the observations produced by different codes are independent given the latent class of the input.
Therefore, if the latent indicators $\{T_{iq}\}$ were observed, the complete-data likelihood is
\[
L_c(\mathbf{p},\boldsymbol{\pi};U,T)
=
\prod_{i=1,\dots,|I_P|}
\prod_{q\in\{0,1\}}
\left(
p_q
\prod_{j\in\Omega_i}
\prod_{a\in\{0,1\}}
\left(\pi^{(j)}_{q a}\right)^{n^{(j)}_{ia}}
\right)^{T_{iq}}.
\]

However, the latent indicators $T_{iq}$ are unobserved.
Marginalizing them out yields the observed-data likelihood
\[
L(\mathbf{p},\boldsymbol{\pi};U)
=
\prod_{i=1,\dots,|I_P|}
\left(
\sum_{q\in\{0,1\}}
p_q
\prod_{j\in\Omega_i}
\prod_{a\in\{0,1\}}
\left(\pi^{(j)}_{q a}\right)^{n^{(j)}_{ia}}
\right).
\]
This likelihood has the same mixture form as in the original Dawid--Skene model, and direct maximization is inconvenient because the summation over latent classes is inside the product over inputs.
We therefore treat $\{T_{iq}\}$ as missing data and optimize the model using EM.

In the E-step, we compute the posterior expectation of each latent indicator
\[
\begin{aligned}
\hat T_{iq}
&\triangleq
E[T_{iq}\mid U] \\
&=
\frac{
\prod_{j\in\Omega_i}\prod_{a\in\{0,1\}}
\left(\pi^{(j)}_{qa}\right)^{n^{(j)}_{ia}} p_q
}{
\sum_{q'=0}^{1} \prod_{j\in\Omega_i}\prod_{a\in\{0,1\}} \left(\pi^{(j)}_{q'a}\right)^{n^{(j)}_{ia}} p_{q'}
}.
\end{aligned}
\]

In the M-step, we maximize the expectation of the complete-data log-likelihood under the posterior distribution computed in the E-step:
\[
\begin{aligned}
E\!\left[\log L_c(\mathbf{p},\boldsymbol{\pi};U,T)\mid T\right]
&= \\
\sum_{i=1,\dots,|I_P|}\sum_{q\in\{0,1\}}
\hat T_{iq}
\Bigl(
\log p_q
&\quad+
\sum_{j\in\Omega_i}\sum_{a\in\{0,1\}}
n^{(j)}_{ia}\log \pi^{(j)}_{qa}
\Bigr).
\end{aligned}
\]

Subject to $\sum_q p_q=1$ and $\sum_a \pi^{(j)}_{qa}=1$ for each $j,q$, maximizing this objective yields the closed-form updates
\[
p_q
\leftarrow
\frac{1}{|I_P|}
\sum_{i=1,\dots,|I_P|}
\hat T_{iq},
\qquad q\in\{0,1\},
\]
and
\[
\pi^{(j)}_{q a}
\leftarrow
\frac{
\sum_{i=1,\dots,|I_P|}
\hat T_{iq}\, n^{(j)}_{ia}
}{
\sum_{i=1,\dots,|I_P|}
\hat T_{iq}
},
\qquad q,a\in\{0,1\},
\]
The EM iterations alternate between these two steps until convergence.
The final estimates $\hat{\mathbf{p}}$ and $\hat{\boldsymbol{\pi}}$ are then used to quantify the reliability of each candidate code.

\subsubsection{\textbf{Reliability Score from Individual Error-Rates}}

Finally, the reliability score $\alpha_j$ of code $C_j$ is derived from its
estimated individual error-rates $\pi^{(j)}$ as
\[
\alpha_j
\triangleq
\hat p_1\,\hat\pi^{(j)}_{1,1}
+
\hat p_0\,\hat\pi^{(j)}_{0,0}
\in [0,1],
\]
which corresponds to the expected probability that $C_j$ produces execution outcomes consistent with the latent execution-consistency states inferred across inputs. For binary decisions, we classify $C_j$ as \textsc{Reliable} if $\alpha_j$ exceeds a task-specific threshold, and as \textsc{Unreliable} otherwise. Among the reliable candidates, we select the code with the highest inferred reliability,
\[
j^{*}=\arg\max_{j:\,\alpha_j\ge\tau}\alpha_j,\qquad
C^{*}=C_{j^{*}},\qquad
\alpha^{*}=\alpha_{j^{*}},
\]
If no candidate satisfies $\alpha_j \ge \tau$, ExeCRE does not select a reference code or add generated reference tests. Self-correction then proceeds from the initial code using only public tests, preventing an unreliable reference code from making a correct initial solution appear to fail.

\subsection{Self-Correction with Reliability Estimation}

Figure~\ref{fig:Self_Correction} illustrates how ExeCRE is integrated into the self-correction pipeline for code generation. Given a programming problem, the LLM produces both a candidate solution and a set of reference codes intended for constructing verification signals. ExeCRE estimates the reliability of these reference codes before they are used downstream.

Reference codes whose reliability scores exceed a predefined threshold are regarded as reliable candidates. Formally, let $C^{*}$ denote the reliable candidate with the highest inferred reliability score $\alpha^{*}$. This selected reference code is then executed on generated small-scale inputs to produce input--output pairs, which serve as reference tests. If no reference code passes the threshold, the verification stage falls back to using only the available public tests.

The reference code is a brute-force implementation obtained by prompting the LLM to solve the problem using direct simulation, explicit enumeration, or similarly straightforward strategies. Prior work has shown that such implementations are often easier for LLMs to generate correctly when efficiency is not a concern~\cite{zhang2023algo}. As a result, brute-force codes tend to exhibit higher correctness compared to optimized solutions, making them suitable as reference implementations. 
However, these codes are typically inefficient and cannot scale to large inputs. We therefore restrict execution to small-scale inputs, where the reference code remains tractable while still providing reliable execution outcomes for both reliability estimation and test construction.

Using the assembled test set, we adopt a test-driven self-correction pipeline following TextGrad~\cite{yuksekgonul2025optimizing}, while avoiding direct LLM-based correctness judgments as feedback signals. When a reliable reference code is available, the resulting test set provides stronger verification signals and supports more effective self-correction iterations. When it is not, ExeCRE prevents unreliable reference tests from entering the loop and reduces harmful corrections caused by misleading execution feedback.

\section{Experimental Setup}

\subsection{Research Questions}

We carry out experiments to answer the following research questions:

\noindent\textbf{RQ1:} \textit{Does ExeCRE improve final performance across models of different scales?}

\noindent\textbf{RQ2:} \textit{To what extent can reliability estimation distinguish reliable code from unreliable code?}

\noindent\textbf{RQ3:} \textit{Does reliability estimation reduce misleading or unnecessary correction signals and improve the stability of iterative self-correction?}

\noindent\textbf{RQ4:} \textit{How robust is ExeCRE's reliability estimation to threshold choices and agreement among wrong candidates, and how sensitive is it to the input construction method?}

We also include an additional applicability study on GSM8K to examine whether the same reliability estimation mechanism can also help code-based mathematical reasoning, and briefly analyze runtime and token cost for practical reference.

\begin{table*}
\centering
\small
\setlength{\tabcolsep}{1.8pt}
\caption{Pass@1 performance across models of different scales, with results further broken down by difficulty level. All numbers are reported as mean $\pm$ standard deviation over five runs.}
\begin{tabular}{lcccccccccccccccc}
\toprule
\multirow{2}{*}{\centering Method}
& \multicolumn{4}{c}{GPT-5.2}
& \multicolumn{4}{c}{DeepSeek-V3.2}
& \multicolumn{4}{c}{Qwen2.5-Coder-32B-Instruct}
& \multicolumn{4}{c}{LLaMA-3.1-8B-Instruct} \\
\cmidrule(lr){2-5} \cmidrule(lr){6-9} \cmidrule(lr){10-13} \cmidrule(lr){14-17}
& All & Easy & Medium & Hard
& All & Easy & Medium & Hard
& All & Easy & Medium & Hard
& All & Easy & Medium & Hard \\
\midrule
Base
& 62.7\std{1.8} & 99.6\std{1.8} & 74.2\std{3.5} & 34.9\std{3.1}
& 44.5\std{1.4} & 91.6\std{1.7} & 48.0\std{3.0} & 16.3\std{2.3}
& 27.4\std{0.6} & 87.1\std{2.6} & 14.9\std{2.1} & 2.9\std{1.0}
& 16.3\std{0.4} & 57.3\std{1.7} & 6.2\std{1.9} & 0.5\std{0.6} \\
\midrule
ConTested
& 55.6\std{1.2} & 74.7\std{1.1} & 67.6\std{2.4} & 37.1\std{1.8}
& 43.6\std{0.5} & 90.2\std{1.1} & 49.1\std{3.0} & 14.4\std{1.2}
& 16.2\std{0.8} & 48.0\std{3.0} & 13.1\std{3.1} & 0.7\std{1.0}
& 7.7\std{0.6} & 27.8\std{1.1} & 2.7\std{0.9} & 0.0\std{0.0} \\
Oracle
& 63.2\std{0.8} & \textbf{100.0}\std{0.0} & 77.0\std{1.3} & 33.7\std{1.5}
& 44.4\std{0.2} & 92.4\std{1.1} & 50.9\std{0.0} & 13.7\std{0.5}
& 26.2\std{0.7} & 80.0\std{1.4} & 18.2\std{2.0} & 2.0\std{0.6}
& 8.2\std{0.2} & 31.1\std{1.1} & 1.8\std{0.2} & 0.0\std{0.0} \\
B4
& 64.5\std{0.6} & \textbf{100.0}\std{0.0} & 77.8\std{1.8} & 36.1\std{0.6}
& 45.6\std{0.6} & 91.1\std{0.0} & 54.6\std{0.2} & 14.6\std{0.0}
& 28.6\std{0.8} & 88.9\std{2.4} & 18.2\std{1.2} & 2.4\std{0.6}
& 10.1\std{1.8} & 36.9\std{6.2} & 3.3\std{0.7} & 0.0\std{0.0} \\
ALGO
& 66.8\std{1.0} & \textbf{100.0}\std{0.0} & 77.8\std{0.6} & 41.2\std{2.0}
& 44.8\std{0.2} & 91.1\std{0.0} & 52.0\std{2.0} & 14.6\std{1.3}
& 28.1\std{1.2} & 85.3\std{1.8} & 17.1\std{2.0} & 4.2\std{1.5}
& 8.1\std{0.4} & 30.7\std{1.7} & 1.8\std{0.2} & 0.0\std{0.0} \\
TextGrad
& 69.9\std{1.2} & 98.2\std{1.7} & 80.7\std{2.5} & 47.1\std{2.3}
& 52.6\std{1.0} & 96.0\std{1.7} & 56.0\std{4.2} & 26.6\std{2.2}
& 31.5\std{1.0} & 86.2\std{4.3} & \textbf{25.5}\std{3.8} & 5.6\std{1.7}
& 19.4\std{0.5} & 67.4\std{1.1} & 7.9\std{1.7} & 0.8\std{0.6}\\
\midrule
\framework
& \textbf{72.8}\std{1.5} & 99.6\std{0.9} & \textbf{84.4}\std{1.9} & \textbf{50.2}\std{0.9}
& \textbf{55.4}\std{1.8} & \textbf{97.8}\std{0.0} & \textbf{62.2}\std{2.7} & \textbf{27.6}\std{3.5}
& \textbf{32.5}\std{1.4} & \textbf{91.1}\std{1.4} & 24.0\std{1.8} & \textbf{6.1}\std{1.8}
& \textbf{21.0}\std{0.9} & \textbf{72.0}\std{3.3} & \textbf{8.7}\std{1.8} & \textbf{1.2}\std{0.8}  \\
\bottomrule
\end{tabular}
\label{tab:pass1-multi-model}
\end{table*}

\subsection{Dataset}

We evaluate the framework on competitive code generation tasks, with experiments conducted on LiveCodeBench.
LiveCodeBench~\cite{zhang2024livecodebench} consists of algorithmic programming problems collected from Codeforces, AtCoder, and LeetCode.
To minimize the potential impact of training-data contamination on evaluation, we use the most recent LiveCodeBench problems available at the time of submission. Specifically, we use LiveCodeBench v6 and include all 182 problems released between 2025-01-01 and 2025-05-01. No later problems were available because the benchmark had not yet been updated beyond May 2025. The complete list of problem identifiers is provided in the archived artifact.

As a small additional applicability study beyond the main code-generation setting, we also evaluate ExeCRE on GSM8K~\cite{cobbe2021training}, using its test set of 1,319 grade-school mathematical word problems. We adopt a program-of-thought (PoT) setting, where the model may either generate executable code or directly produce the answer.

\subsection{Evaluation Metrics}

For LiveCodeBench, we report final Pass@1 under the default evaluation setting.
Selected brute-force codes are used as reference codes to guide self-correction.
To evaluate the quality of these selected codes, we also report F1, Precision, and Recall under a semantic-only setting, where Time Limit Exceeded (TLE) cases are treated as correct because computational efficiency is not a concern for brute-force reference code.
For GSM8K, we report answer accuracy, defined as the proportion of problems whose final numerical answers exactly match the ground truth.

\subsection{Compared Techniques}

We compare \framework on LiveCodeBench against several representative baselines.
\textbf{Base} directly generates a single solution using the standard LiveCodeBench prompt.

\textbf{ConTested}~\cite{dong2025contested} performs consistency-based, test-driven self-correction by co-evolving code and tests. 
For fair comparison, we use the same model for both test generation and refinement, rather than relying on stronger external models or human feedback as in the original setup.
\textbf{Oracle-Guided}~\cite{fan2024oracleguided} performs program selection using LLM-guided oracle judgments on distinguishing inputs.
\textbf{B4}~\cite{chen2024b4} performs Bayesian solution selection under noisy LLM-generated test signals.
\textbf{ALGO}~\cite{zhang2023algo} uses brute-force reference codes to generate oracle tests and guide code selection within existing test-driven frameworks, such as CodeT~\cite{chen2023codet}.
\textbf{TextGrad}~\cite{yuksekgonul2025optimizing} performs LLM-based iterative self-correction.
For GSM8K, we compare with the program-of-thought (PoT) baseline, as our goal is to evaluate whether ExeCRE can effectively estimate the reliability of generated code in mathematical reasoning settings, rather than to comprehensively benchmark against all reasoning methods.

\subsection{Implementation Details}

We evaluate GPT-5.2~\cite{openai_gpt52_2025},
DeepSeek-V3.2~\cite{liu2025deepseek},
Qwen2.5-Coder-32B-Instruct~\cite{hui2024qwen2_5_coder},
and LLaMA-3.1-8B-Instruct~\cite{grattafiori2024llama}
on LiveCodeBench. GPT-5.2, Qwen2.5-Coder-32B-Instruct, and
LLaMA-3.1-8B-Instruct have reported knowledge cutoffs of
2025-08, 2024-06, and 2023-12, respectively. DeepSeek-V3.2
was released on December 1, 2025, but no precise knowledge
cutoff has been disclosed. We additionally evaluate
Qwen2.5-Coder-32B-Instruct and LLaMA-3.1-8B-Instruct on GSM8K.

For Dawid–Skene aggregation, we sample 10 reference code candidates and 5 schemas for execution input construction. Reference codes are generated with temperature set to 1.0 and top-$p$ to 0.99 to encourage diversity. During execution outcome matrix construction, we generate 300 inputs for each code using our schema-driven execution input generation strategy, resulting in execution output matrices of size approximately $10 \times 300$.
We use greedy decoding (temperature=0.0) in self-correction to ensure deterministic behavior. All baselines involving multi-round self-correction are run for 20 rounds.

During test construction, we adopt 0.95 as the reliability score threshold. When evaluating code in self-correction, we constrain the test suite to execute within 1 minute under a single-threaded setting; in practice, this allows us to generate approximately 2,000 test cases.
The reported results are averaged over 5 runs, and we report the mean and standard deviation (mean ± std) to account for stochasticity.

\section{Experiment Results}

\subsection{RQ1: Overall Performance Across Model Scales}
\label{sec:rq1}

We compare \framework against representative self-correction and selection baselines on four models of varying scales. As shown in Table~\ref{tab:pass1-multi-model}, \framework achieves the best overall Pass@1 on all four models.
Cutoff cleanliness varies across models, as reported knowledge cutoffs of Qwen2.5-Coder-32B-Instruct and LLaMA-3.1-8B-Instruct precede the entire evaluated window.
In contrast, the cutoff of GPT-5.2 postdates the entire window, and no cutoff-clean subset was available for this model in the most recent LiveCodeBench release available at the time of our evaluation.
For DeepSeek-V3.2, the developer does not disclose a precise knowledge cutoff, so the cutoff cleanliness of its results and the availability of a strictly cutoff-clean subset cannot be verified.
We therefore use the GPT-5.2 and DeepSeek-V3.2 results for comparison but do not consider them contamination-free evidence.

The gains are most evident on Medium and Hard problems, while all methods perform similarly on Easy problems due to near-saturated accuracy. This pattern suggests that \framework is particularly beneficial in more challenging settings, where unreliable reference codes are more likely to mislead downstream correction.

The same trend is observed on the two models whose knowledge cutoffs precede the benchmark period. On Qwen2.5-Coder-32B-Instruct, \framework increases Pass@1 from 31.5 to 32.5 and reduces misleading feedback from 50.4 to 7.4. On LLaMA-3.1-8B-Instruct, Pass@1 increases from 19.4 to 21.0, while misleading feedback decreases from 30.0 to 1.2. These results provide cutoff-clean support for the main qualitative finding that reliability-aware selection improves self-correction.

The overall results in Table~\ref{tab:pass1-multi-model} show that reliability-aware selection is beneficial in general, but they do not by themselves explain how well \framework distinguishes semantically reliable code candidates or how this advantage affects the correction process. We therefore next examine these two aspects through semantic-level reliability identification in RQ2 and misleading-feedback reduction together with iterative stability in RQ3.

\subsection{RQ2: Effectiveness of Reliability Estimation}
\label{sec:rq2}

We evaluate whether reliability estimation from execution consistency can effectively distinguish correct from incorrect codes at the semantic level. 
Each problem is treated as one evaluation instance. A positive prediction means that the candidate with the highest score is adopted because its reliability score reaches $\alpha=0.95$, while the ground truth label is positive only when every hidden test is Accepted or TLE and none fails.
Specifically, we consider a semantic-only setting where Time Limit Exceeded is treated as correct, and compare predicted correctness against ground-truth labels using Precision, Recall, and F1.
In the context of self-correction, Precision reflects the risk of introducing misleading feedback, while Recall measures the ability to preserve correct signals. We therefore use F1 as the primary metric to capture the trade-off between the two.
Otherwise, ExeCRE falls back to public tests. Fallback is the percentage of problems where no reference program is adopted.

\subsubsection{Dawid--Skene vs Majority Voting: An Example}
We use a simple example to show how Dawid--Skene can distinguish reliable codes more clearly when majority voting assigns similar scores to correct and incorrect candidates. 
Consider a coupon rule that is valid only when \texttt{subtotal >= 25} and \texttt{count >= 3}. Codes A, B, and C implement the rule correctly. Codes D and E use \texttt{subtotal > 25}, while codes F and G use \texttt{count > 3}.

\begin{table}[ht]
    \centering
    \small
    \setlength{\tabcolsep}{4pt}
    \caption{A case where agreement among wrong codes misleads majority voting. T and F indicate that the coupon is valid and invalid, respectively.}
    \label{tab:ds-voting-example}
    \begin{tabular}{lcccccccc}
        \toprule
        Input & A & B & C & D & E & F & G & Voting \\
        \midrule
        subtotal=25, count=5 & T & T & T & F & F & T & T & T (5:2) \\
        subtotal=40, count=3 & T & T & T & T & T & F & F & T (5:2)\\
        subtotal=25, count=3 & T & T & T & F & F & F & F & F (3:4)\\
        \bottomrule
    \end{tabular}
\end{table}

In the first two inputs, only one type of bug is exposed, and the majority voting decision remains correct. In the final input, both bug types are triggered, causing four incorrect programs to agree on F causes majority voting to make an incorrect decision.
Across the three inputs, ExeCRE-Voting assigns A--C a score of $13/21$ and D--G a score of $11/21$. Since both scores remain below the 0.95 adoption threshold, ExeCRE-Voting falls back. 
Dawid--Skene instead jointly estimates per-code error-rates by maximizing the likelihood of the full consistency matrix. Applying Dawid--Skene to this matrix assigns substantially higher reliability to A--C than to D--G, allowing it to adopt one of the correct codes.

\subsubsection{Comparison with Reliability Identification Methods}

Table~\ref{tab:deepseek-metrics} reports detailed results on DeepSeek-V3.2.
\textbf{Random} randomly decides whether to adopt the selected reference program or fall back, without examining the problem or code.
\textbf{LLM Judge} directly asks the model to assess the code correctness.
\textbf{LLM Analyze} follows the TextGrad-style prompting strategy~\cite{yuksekgonul2025optimizing}, where the model first generates an analysis before making a judgment.
\textbf{CodeJudge}~\cite{tong2024codejudge} uses its Analyze then Summarize method, first analyzing the candidate program and then producing a YES or NO judgment from the analysis.
\textbf{SLM Judge}~\cite{crupi2026improving} uses a Qwen2.5 Coder 3B model fine-tuned for binary code correctness classification. We score each candidate using the probability of label 1 normalized over labels 0 and 1, and adopt the highest scored candidate only when its score exceeds 0.06, giving SLM Judge a fallback rate close to ExeCRE's and similar adoption coverage. None of the methods has access to hidden tests.
We further compare with \vtsub and \framework, which estimate reliability based on execution consistency using voting and EM-based inference, respectively.

\begin{table}[ht]
    \centering
    \small
    \setlength{\tabcolsep}{4pt}
    \caption{Reliability identification performance on DeepSeek-V3.2. Both \framework and \vtsub use a reliability threshold of 0.95.}
    \begin{tabular}{lcccc}
        \toprule
        Method & F1 (\%) & Precision (\%) & Recall (\%) & Fallback (\%) \\
        \midrule
        Random              & 56.06 & 52.22 & 60.50 & 39.50 \\
        LLM Judge           & 60.49 & 94.23 & 44.55 & 71.43 \\
        LLM Analyze         & 61.82 & 92.73 & 46.36 & 69.78 \\
        CodeJudge           & 62.58 & 94.44 & 46.79 & 69.49 \\
        SLM Judge           & 72.64 & 82.95 & 64.60 & 50.28 \\
        \midrule
        \vtsub              & 70.83 & 87.18 & 59.65 & 55.37 \\
        \framework          & \textbf{76.62} & 86.52 & 68.75 & 48.55 \\
        \bottomrule
    \end{tabular}
    \label{tab:deepseek-metrics}
\end{table}

As shown in Table~\ref{tab:deepseek-metrics}, both \framework variants significantly outperform all baselines in terms of F1, demonstrating stronger semantic discrimination ability. LLM-based methods achieve high precision but suffer from low recall, indicating conservative but incomplete identification of correct codes. In contrast, \framework achieves a more balanced trade-off, reducing misleading feedback while retaining substantially more correct signals. This gap between ExeCRE and ExeCRE-Voting suggests that simple output agreement is not sufficient for reliability identification. By modeling code-specific error tendencies, the EM-based estimator can retain more semantically correct candidates while still controlling misleading signals.
\framework achieves the highest F1. SLM Judge also performs well, but has slightly lower precision and recall.

In terms of LLM call cost beyond shared candidate generation, Random requires no model calls, and SLM Judge uses only a fine tuned local model. Each remaining selector requires fewer than five LLM calls per problem.
In terms of runtime under single threaded execution without parallel judge calls, CodeJudge completes within 15 seconds per problem, while ExeCRE Voting and ExeCRE complete within 30 seconds. The remaining methods complete within 5 seconds, with most of their runtime coming from the latency of a single LLM call.

\subsubsection{Reliability Identification Across Models}

\begin{table}[ht]
    \centering
    \small
    \setlength{\tabcolsep}{4pt}
    \caption{RQ2 reliability identification results across models.}
    \label{tab:rq2-across-models}
    \begin{tabular}{lccc}
        \toprule
        Model & F1 (\%) & Precision (\%) & Recall (\%) \\
        \midrule
        GPT-5.2    & 78.12 & 88.50 & 69.93 \\
        DS-V3.2    & 76.62 & 86.52 & 68.75 \\
        Qwen-32B   & 74.50 & 69.10 & 80.90 \\
        LLaMA-8B   & 55.60 & 50.10 & 62.50 \\
        \bottomrule
    \end{tabular}
\end{table}

To check whether the RQ2 result is specific to DeepSeek-V3.2, Table~\ref{tab:rq2-across-models} reports ExeCRE's reliability identification results across all four evaluated models under the same experimental settings.
The results show that reliability identification is not specific to a single generator. ExeCRE reaches similar F1 on GPT-5.2, DeepSeek-V3.2, and Qwen-32B, and the LLaMA-8B row provides the same evaluation on a smaller model. Together with Table~\ref{tab:deepseek-metrics}, these results support that execution consistency provides a useful reliability signal across model families.

\begin{table}[ht]
    \centering
    \small
    \setlength{\tabcolsep}{4pt}
    \caption{Number of problems on the 182-problem LiveCodeBench set for which self-correction methods produce feedback signals for already correct codes. Results are reported as mean $\pm$ standard deviation over five runs.}
    \begin{tabular}{lcccc}
        \toprule
        Method & GPT-5.2 & DS-V3.2 & Qwen-32B & LLaMA-8B \\
        \midrule
        TextGrad  & 113.20$\pm$3.06 & 81.20$\pm$2.99 & 50.40$\pm$0.80 & 30.00$\pm$0.82 \\
        ConTested & 104.60$\pm$1.36 & 72.20$\pm$0.40 & 48.40$\pm$0.80 & 10.00$\pm$1.44 \\
        \midrule
        \framework & 14.00$\pm$1.67 & 11.60$\pm$1.74 & 7.40$\pm$0.80 & 1.20$\pm$0.75 \\
        \bottomrule
    \end{tabular}
    \label{tab:misleading-feedback}
\end{table}

\subsection{RQ3: Mitigating Misleading Feedback and Improving Iterative Stability}
\label{sec:rq3}

\subsubsection{Reduction of Misleading Feedback}

Table~\ref{tab:misleading-feedback} reports the number of problems where a self-correction method produces feedback signals for already correct codes, thus triggering unnecessary correction steps.

Both TextGrad and ConTested generate such misleading signals on a large number of problems across models, whereas \framework consistently reduces this number by a substantial margin.
Some misleading feedback may preserve correctness, but it still adds cost and instability. Reliability estimation reduces such unnecessary interventions.
For example, in LeetCode 3743, the initial solution passed both public and hidden tests, but tests generated by a low reliability reference code reported a Rand failure on it. ExeCRE avoids this misleading feedback by excluding tests generated from reference codes whose reliability score is below 0.95.

\subsubsection{Iterative Self-Correction Stability}

\begin{figure}
    \centering
    \includegraphics[width=\columnwidth]{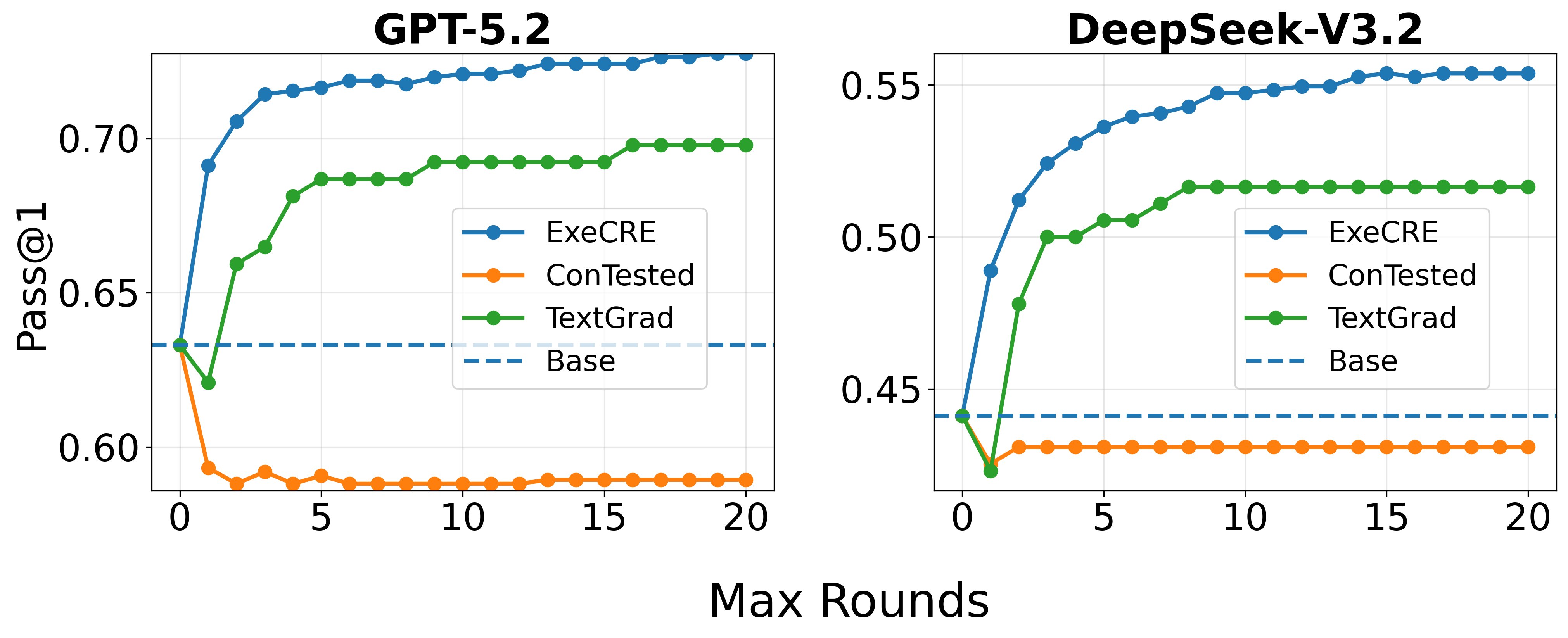}
    \caption{Pass@1 over self-correction iterations.}
    \label{fig:RH_RQ5}
\end{figure}

Figure~\ref{fig:RH_RQ5} further examines the iterative behavior of different self-correction methods. A consistent trend is observed across both models. \framework improves rapidly in the early rounds and ultimately achieves the best final Pass@1. On GPT-5.2, \framework continues to improve in later iterations rather than saturating early, and a similar steady increase is observed on DeepSeek-V3.2.

In contrast, TextGrad yields only limited early gains and quickly saturates, while ConTested remains below the base and stays nearly flat across iterations, suggesting ineffective or misleading correction signals. One possible reason is that, in our implementation, the same latest model is used for both test generation and test modification in ConTested, whereas the original setting relies on a stronger model or human involvement for test refinement.
Notably, \framework does not exhibit early-round Pass@1 degradation on either model, which further supports the effectiveness of its reliability-guided mechanism in suppressing misleading corrections and stabilizing iterative refinement.

\subsection{RQ4: Robustness of Reliability Estimation and Input Construction}
\label{sec:rq4}

\subsubsection{Sensitivity to the Reliability Threshold}
\begin{figure}
    \centering
    \includegraphics[width=\columnwidth]{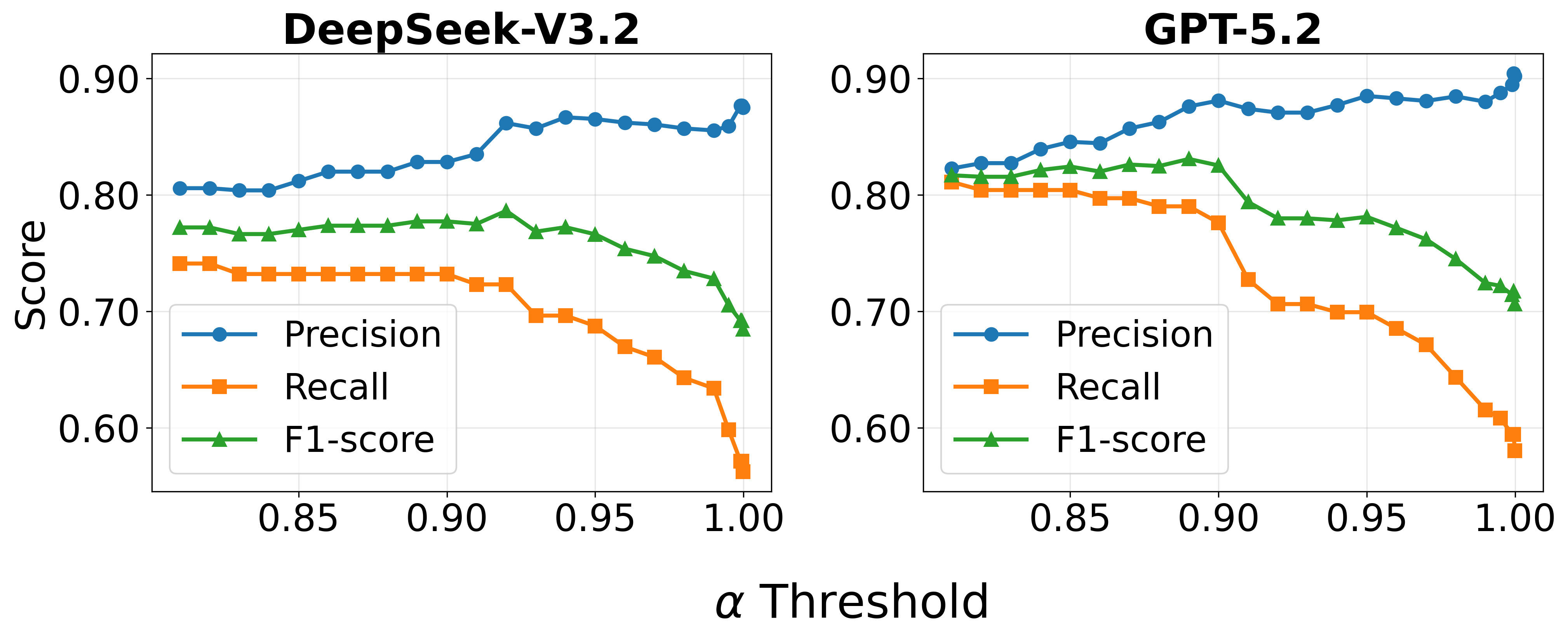}
    \caption{Precision, Recall, and F1 of EM under different reliability thresholds.}
    \label{fig:RH_RQ4}
\end{figure}

Figure~\ref{fig:RH_RQ4} shows how Precision, Recall, and F1 vary with the reliability threshold $\alpha$ on GPT-5.2 and DeepSeek-V3.2. As $\alpha$ increases, Precision improves and Recall decreases, while F1 remains stable over a broad range of thresholds. This suggests that the EM based reliability estimator is not sensitive to careful threshold tuning.

\subsubsection{Robustness to Input Construction}

The pipeline constructs execution inputs from extracted schemas. Based on human verification, schema extraction succeeds for most problems, reaching 96.7\% on GPT-5.2, 95.6\% on LLaMA-3.1-8B-Instruct, 92.9\% on DeepSeek-V3.2, and 81.3\% on Qwen2.5-Coder-32B-Instruct. When extraction is wrong, the resulting invalid or inconsistent executions usually lead to low reliability scores rather than confident adoption.

\begin{table}[ht]
    \centering
    \small
    \setlength{\tabcolsep}{4pt}
    \caption{Reliability identification with different input generators on 65 LiveCodeBench function-call problems.}
    \begin{tabular}{lccc}
        \toprule
        Input source & F1 & Precision & Recall \\
        \midrule
        Schema(Ours) & 0.791 & \textbf{0.872} & 0.723 \\
        Fuzzing(Oracle-Guided) & 0.795 & 0.795 & 0.795 \\
        LLM generate(CodeT) & \textbf{0.813} & 0.804 & \textbf{0.822} \\
        \bottomrule
    \end{tabular}
    \label{tab:input-generator-ablation}
\end{table}

To test whether the reliability estimator depends on schema-based input construction, we replace our schema-based input generator with Oracle-Guided fuzzing~\cite{fan2024oracleguided} and LLM-generated inputs from CodeT~\cite{chen2023codet}. As shown in Table~\ref{tab:input-generator-ablation}, Oracle-Guided fuzzing is implemented for typed function-call problems and does not directly support stdin-style input generation, so this ablation uses the 65 LiveCodeBench problems with function-call interfaces. All three input sources achieve F1 scores between 0.79 and 0.81, suggesting that the reliability estimator is not tied to schema-based input construction on this subset.

\subsubsection{Robustness to Agreement Among Wrong Candidates}

Dawid--Skene assumes conditional independence, but reference codes sampled from the same LLM may share mistakes. We therefore test whether stronger agreement among wrong candidates increases incorrect adoption. Incorrect adoption refers to adopting a reference code that is semantically incorrect. Wrong groups are formed by identical wrong outputs across generated inputs. Fewer wrong groups mean that more wrong candidates behave identically, indicating stronger dependence and a larger departure from conditional independence.

\begin{table}[ht]
    \centering
    \small
    \setlength{\tabcolsep}{5pt}
    \caption{Incorrect adoption by number of wrong output groups. Results are pooled over four models. Share is over 233 mixed model problem instances, and Wrong codes reports the bucket average.}
    \label{tab:wrong-output-groups}
    \begin{tabular}{lccc}
        \toprule
        Wrong groups & Share (\%) & Wrong codes & Incorrect adoption (\%) \\
        \midrule
        1  & 25.3 & 1.9 & 10.2 \\
        2  & 17.6 & 3.8 & 7.3 \\
        3  & 13.7 & 5.1 & 3.1 \\
        4  & 9.0  & 5.9 & 14.3 \\
        5+ & 34.3 & 7.8 & 10.0 \\
        \bottomrule
    \end{tabular}
\end{table}

Table~\ref{tab:wrong-output-groups} reports results pooled across the four evaluated models. Each instance has at least one correct and one wrong reference code. If such dependence undermined Dawid--Skene, incorrect adoption should increase as the number of wrong groups decreases. Instead, incorrect adoption remains between 3.1\% and 14.3\% across all group counts, including 10.2\% when there is one wrong group. This supports the robustness of the estimator under the observed dependence among errors from the same LLM.

\subsection{Additional Applicability Study on Code-Based Mathematical Reasoning}
\label{sec:rq5}

\begin{figure}[htbp]
    \centering
    \includegraphics[width=\columnwidth]{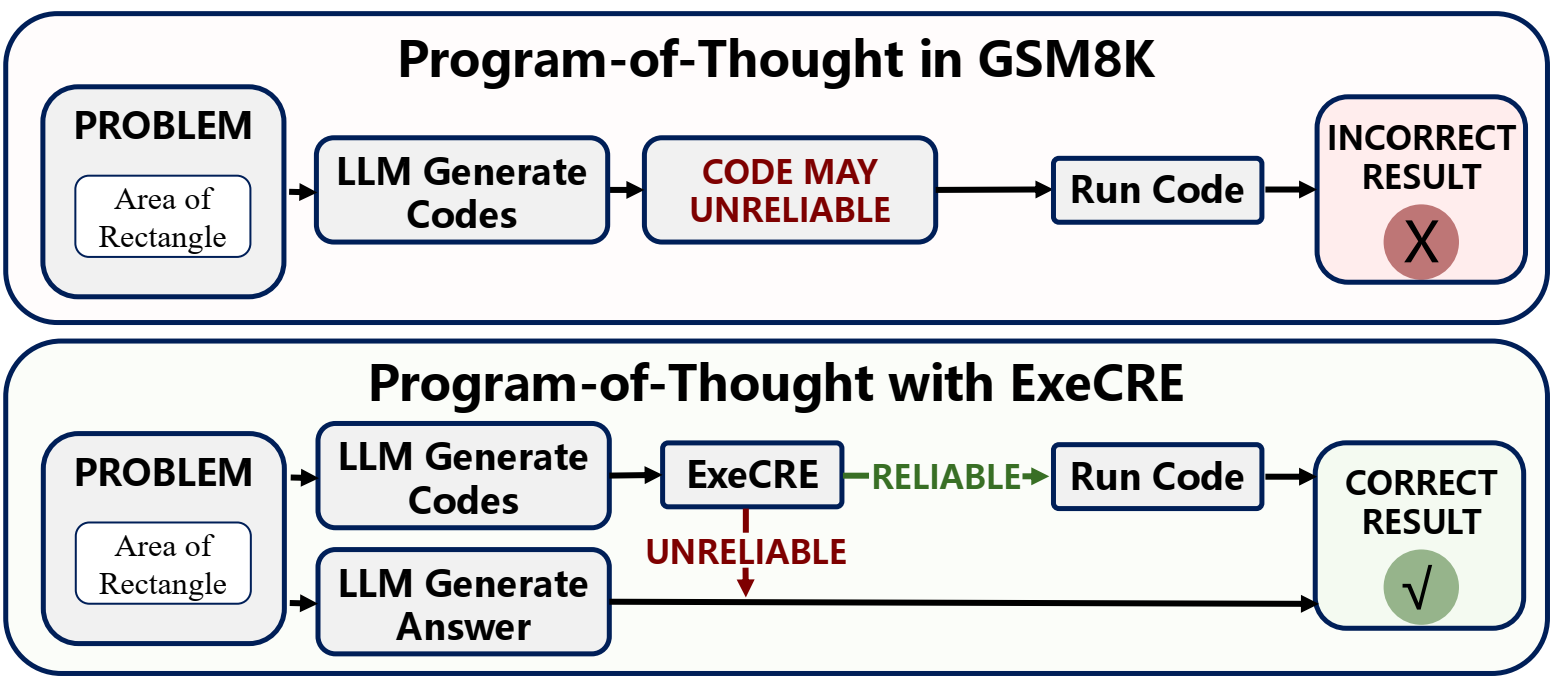}
    \caption{Comparison between the original PoT pipeline and PoT enhanced with \framework in the additional GSM8K study.}
    \label{fig:pot_math}
\end{figure}

As a small additional applicability study, we further examine whether the same reliability estimation mechanism can also help in code-based mathematical reasoning. We evaluate \framework on GSM8K under a program-of-thought (PoT) setting, where the model solves each problem by generating executable code and deriving the final answer from execution~\cite{chen2022program}. In this setting, \framework is not used to construct verification signals for iterative correction. Instead, it serves as a reliability gate for generated code. Only candidates whose reliability is sufficiently supported by execution evidence are executed to produce the final answer. Otherwise, the system falls back to the direct-answer path. Figure~\ref{fig:pot_math} shows the overall pipeline of PoT enhanced with \framework in this setting, together with an illustrative GSM8K example (``Area of Rectangle'').

Compared with general code generation tasks, execution input construction in GSM8K is also simpler. We construct execution inputs by changing the explicit numeric parameters in the generated code within bounded ranges while preserving the original problem structure. This perturbation-based strategy yields randomized inputs that remain close to the original instance while still exposing behavioral differences among candidate code solutions, without requiring complex schema extraction or dependency parsing.

We focus on two relatively smaller models, Qwen2.5-Coder-32B-Instruct and LLaMA-3.1-8B-Instruct, because GSM8K is already close to saturation for stronger frontier models, making it less suitable for analyzing the effect of reliability estimation. Since our goal here is only to examine whether reliability estimation can also help when reasoning relies on generated code, we mainly compare \framework with PoT, rather than strategies that do not rely on code generation.

The results suggest that ExeCRE can also provide improvements over PoT in this setting. On Qwen2.5-Coder-32B-Instruct, accuracy increases from 90.28\% to 92.96\%; on LLaMA-3.1-8B-Instruct, it improves from 75.16\% to 78.85\%. These results provide additional evidence that reliability estimation can also be useful in mathematical reasoning settings when the model relies on generated code.
Using the same global reliability threshold $\alpha = 0.95$, \framework achieves 94.93\% precision, 53.49\% recall, and 68.42 F1 in identifying reliable code paths. This conservative behavior matches the reasoning setting, where executing unreliable code can directly determine the final answer. The gains therefore come from better deciding when generated code should be trusted, while the remaining cases fall back to direct answers.

\subsection{Runtime Analysis}
\label{sec:cost}

Under a practical multithreaded setting, we sample 20 tasks and repeat the experiment five times, yielding 100 runs with batched DeepSeek API calls. Each task uses 10 candidate codes, five schemas, and 300 inputs per schema. 
ExeCRE takes 26.10 ± 4.37 seconds per run. API calls average 6.40 seconds, while Dawid–Skene estimation takes less than 3 seconds, indicating that evidence collection dominates. 
Parallel execution, batched generation, and smaller candidate or input budgets can reduce this overhead. In costly correction workflows, early filtering may also avoid unnecessary refinement and offset part of the runtime cost.

\section{Related Work}

Generated code serves as both a final answer and an executable component in downstream workflows.
Prior work improves reliability through refinement, candidate selection, and execution or testing~\cite{yuksekgonul2025optimizing,shinn2023reflexion,wang2023selfconsistency,dong2025contested,fan2024oracleguided}.
Other studies evaluate programs using LLM judgments and execution feedback~\cite{yuksekgonul2025optimizing,chen2023codet,chen2024b4,chen2025revisit,he2024cocost}.

\subsection{LLM-as-Judge and Self-Correction}
Several prior works improve generated codes using LLM-based feedback.
Reflexion~\cite{shinn2023reflexion} and Self-Refine~\cite{madaan2024selfrefine}
enable iterative refinement by prompting models to reflect on their own
failures, while Revisit Self-Debugging~\cite{chen2025revisit} further studies
self-debugging with self-generated tests and execution traces. Cycle~\cite{ding2024cycle} studies self-refinement using available feedback and test-suite execution results.
More generally, approaches such as Tree of Thoughts~\cite{yao2023tree},
multi-LLM debate~\cite{du2024improving,estornell2024multi}, MapCoder~\cite{islam2024mapcoder}, and
TextGrad~\cite{yuksekgonul2025optimizing} use structured reasoning,
comparison, or gradient-like textual feedback to guide generation. Dong et al.~\cite{dong2024self} further study self-collaboration for code generation, where multiple LLM agents with specialized roles such as analyst, coder, and tester collaboratively solve complex programming tasks.
CodeTree~\cite{li2025codetree} further introduces agent-guided tree search that combines execution-based and LLM-generated feedback to refine and rank candidate solutions.
While effective, these methods typically rely on feedback whose quality is not
explicitly modeled, which may introduce unstable or misleading correction
signals.

\subsection{Consistency and Candidate Selection}
Another line of work improves generation by sampling multiple candidates and
selecting outputs based on agreement or consistency. Related work also studies reviewer or reranker-based selection for code generation~\cite{zhang2023codereviewer,rafailov2023direct}. Self-Consistency~\cite{wang2023selfconsistency} aggregates multiple reasoning
paths via majority voting, while subsequent extensions explore more diverse
perspectives for consistency-based selection, such as
Multi-Perspective Self-Consistency~\cite{huang2024enhancing}.
CodeChain~\cite{le2023codechain} further extends this idea to modular and
iterative code construction.
Related work also considers explicit similarity-based voting among codes.
For example, More Agents Is All You Need~\cite{li2024more} performs majority
selection based on pairwise BLEU scores between generated codes.
Prior work has used agreement, similarity, and behavioral disagreement to assess generated code. Incoherence~\cite{valentin2026incoherence} is an oracle-free measure of incorrectness. For a programming task, pointwise incoherence is the probability that two independently sampled programs produce different outputs on an input drawn from an input generator. This disagreement probability yields a provable lower bound on the pointwise error of the code generation system.
ExeCRE uses execution consistency for a different self-correction decision. It estimates the reliability of each candidate reference code, selects a reference code only when its score exceeds the threshold, and otherwise does not add generated reference tests to the correction loop.

\subsection{Mitigating Unreliable Feedback Signals}
Several studies seek to reduce the impact of unreliable code, tests, or
verification signals on downstream code selection and refinement.
CodeT~\cite{chen2023codet} evaluates candidate codes using model-generated tests and execution agreement.
ALGO~\cite{zhang2023algo} further introduces brute-force reference codes to generate tests and integrates this signal into test-driven frameworks such as CodeT to improve selection and correction performance.
ConTested~\cite{dong2025contested} enhances self-correction by iteratively
repairing unreliable model-generated tests, thereby reducing misleading
feedback during refinement.
B4~\cite{chen2024b4} studies the challenging setting where both candidate
solutions and generated tests may be plausible but unreliable, and formulates solution selection under such noisy signals in a Bayesian framework.
Related efforts such as ROCODE~\cite{jiang2024rocode} incorporate program analysis or structured search to reduce syntactic or compilation errors.

These methods improve robustness to imperfect feedback signals, but they mainly use such signals for direct selection, filtering, or repair, while the reliability of the executable code component itself is less explicitly modeled before downstream use.
ExeCRE applies reliability filtering before tests generated from a reference code are added to the correction loop.
Mu et al.~\cite{mu2024clarifygpt} propose ClarifyGPT, a framework that detects ambiguous requirements and improves code generation by asking targeted clarifying questions before final code synthesis. Similarly, Fakhoury et al.~\cite{fakhoury2024llm} propose TICODER, an interactive workflow that uses tests to guide intent clarification between users and LLMs, thereby improving code generation under ambiguous natural language specifications.
Related work also explores generating test oracles directly from source code and documentation. Tratto~\cite{molinelli2025tratto}, for instance, derives axiomatic test oracles from Java methods and Javadoc using a neuro-symbolic token-by-token generation procedure.

\subsection{Mathematical Reasoning Using Code}

Generated code can also serve as an executable intermediate for reasoning.
Chain of Code~\cite{li2023chain} combines interpreter execution with
language model emulation, while PAL~\cite{gao2023pal},
Program-of-Thought~\cite{chen2022program}, and
TORA~\cite{gou2023tora} generate code or invoke tools for mathematical reasoning on benchmarks such as GSM8K~\cite{cobbe2021training}.
Recent work further integrates code execution into the reasoning process~\cite{jung2025code,mai2025agentic}.
Because unreliable code can directly determine final answers, these settings motivate reliability-aware use of executable components.

\section{Threats to Validity}
Regarding internal validity, performance may be affected by prompt format, candidate sampling, correction budget, execution limits, and input construction. We use consistent settings where possible and report averages across multiple runs. These choices may still affect absolute performance, but are unlikely to change the overall trends.

Benchmark contamination is another internal validity threat. The evaluated LiveCodeBench window is cutoff-clean for Qwen2.5-Coder-32B-Instruct and LLaMA-3.1-8B-Instruct, but not for GPT-5.2. For DeepSeek-V3.2, cutoff cleanliness cannot be determined because its precise knowledge cutoff is not disclosed. We treat the GPT-5.2 and DeepSeek-V3.2 results as comparative evidence only. Using the same problems for all methods controls the comparison but does not guarantee that potential contamination affects them equally.
Dawid--Skene assumes candidate outputs are conditionally independent given the latent state. This assumption is approximate because reference codes sampled from the same LLM can share bugs. Our wrong output group analysis shows that incorrect adoption remains limited across observed group counts. Still, ExeCRE can adopt an incorrect reference code when many wrong candidates share a bug that generated inputs do not expose.

Regarding external validity, our evaluation does not directly cover repository-level tasks. The main difference in extending ExeCRE to repository-level settings lies in the form of generated inputs. The input-generator ablation in Table~\ref{tab:input-generator-ablation} provides preliminary evidence that the reliability identification pipeline can remain effective when the input generator is replaced, although direct evaluation on repository-level tasks is still needed. Such an extension also requires candidate programs to be executed repeatedly within a reasonable amount of time and with manageable computational and system resources. Evaluating ExeCRE on stateful repository tasks and longer coding-agent workflows remains future work.

\section{Conclusion}

In this work, we study how execution consistency can be leveraged to estimate code reliability and support self-correction in LLM-based code generation.
We propose ExeCRE, an execution-consistency guided reliability estimation framework that infers the trustworthiness of candidate codes from their execution behaviors over large and diverse constructed inputs.
By executing multiple candidate codes and modeling their agreement patterns across inputs, ExeCRE derives reliability scores for candidate codes using a principled aggregation approach.
Experimental results show that this reliability signal helps identify more reliable reference codes and supports more stable iterative self-correction in code generation.
We also include a small additional study on GSM8K, whose results suggest that the same mechanism may be useful for code-based mathematical reasoning when final answers depend on generated code.
Overall, our findings highlight the practical value of execution consistency as a signal for reliability estimation.

\begin{acks}
This work was supported by the National Natural Science Foundation of China (No. U2433212), the China Postdoctoral Science Foundation (Nos. GZB20250951, 2026T191088, and 2026M794685), in part by the Fundamental Research Funds for the Central Universities, and in part by the State Key Laboratory of Complex \& Critical Software Environment.
\end{acks}

\section*{Data Availability Statement}
Artifacts are archived at Zenodo~\cite{dong2026execreartifact}.
The actively maintained repository is available at \url{https://github.com/moyi-dong/ExeCRE}.

\makeatletter
\let\ASEoriglbibitem\@lbibitem
\def\@lbibitem[#1]#2{%
  \def\ASEthiskey{#2}%
  \def\ASEbreakkey{valentin2026incoherence}%
  \ifx\ASEthiskey\ASEbreakkey
    \newpage
  \fi
  \ASEoriglbibitem[#1]{#2}%
}
\makeatother
\bibliographystyle{ACM-Reference-Format}
\bibliography{references}

\end{document}